\documentclass[12pt,a4paper]{article}
\usepackage[english]{babel}
\usepackage[T2A]{fontenc}
\usepackage[cp1251]{inputenc}
\usepackage{amsmath}
\usepackage{graphicx}
\usepackage{amssymb}
\usepackage{color}
\usepackage{amsfonts}
\usepackage{wrapfig}
\usepackage{caption}

\usepackage{citehack,cite}
\usepackage{hyperref}

\newcommand{\non}{\nonumber \\}
\newcommand{\ve}[1]{{\bf #1}}
\newcommand{\be}{\begin{equation}}
\newcommand{\ee}{\end{equation}}
\newcommand{\bea}{\begin{eqnarray}}
\newcommand{\eea}{\end{eqnarray}}

\begin{document}

\begin{center}
{\bf INFLUENCE OF ATTRACTIVE PARTS OF INTERACTION POTENTIALS ON
CRITICAL POINT PARAMETERS}
\end{center}

\begin{center}
{\sc I.V. Pylyuk\footnote{e-mail: piv@icmp.lviv.ua}, O.A. Dobush,
M.P. Kozlovskii, R.V. Romanik, M.A. Shpot}
\end{center}

\begin{center}
{\it Yukhnovskii Institute for Condensed Matter Physics
of the National Academy of Sciences of Ukraine,
1~Svientsitskii Str., 79011 Lviv, Ukraine}
\end{center}

\vspace{0.5cm}

\begin{center}
	{\small\bf{Abstract}}
\end{center}

{\small
We investigate how microscopic features of interparticle potentials influence macroscopic critical point parameters. Our analytical calculations are based on the cell model for continuous many-particle systems.
We explore two types of pair interactions described by the Morse potential and a Curie-Weiss-type potential.
For Morse fluids, we present numerical results obtained with microscopic parameters corresponding to the alkali metals sodium (Na) and potassium (K).
The calculated dimensionless critical point parameters for liquid Na and K, expressed in dimensional units, allow for direct comparison with available experimental and simulation data.
For the Curie-Weiss cell model with competing interactions, which exhibits a sequence of first-order phase transitions, we examine the critical parameters for the first three critical points.
We analyze our results by varying the attractive part of the Morse potential and the Curie-Weiss attraction strength, providing insights into how these microscopic characteristics change critical point coordinates.
}

\vspace{0.5cm}



\section{Introduction}
\label{sec1}

The study of critical behavior in many-particle systems, particularly the behavior of fluids near the liquid-gas critical point, remains one of the central themes in statistical physics \cite{amo191,bsa110,hm113,v118}. Of particular interest is the problem of establishing connections between microscopic characteristics of interparticle interactions and macroscopic parameters of the critical point, such as temperature, density, and pressure. Studying this problem deepens our understanding of phase transitions and supports the development of effective models for real fluids. Also, the influence of interactions receives considerable attention in the interpretation of experimental data. For example, the enhancement of solute-solvent interactions caused by the influence of ionic impurities (as, for example, the NaCl salt) on aqueous solutions of organic molecules leads to an increase in the size of water-solute clusters and results in a reduction of the asymptotic critical region and a decrease in the critical viscosity amplitude \cite{o197}.

Given the important role of interparticle interactions in shaping the critical properties, it is reasonable to focus specifically on the attractive part of the potential, since changes in this component can influence the position of the critical point.

The aim of this work is to evaluate the influence of parameters of the attractive part of  potential on the critical point characteristics in a continuous model system. To this end, we consider both the classical Morse potential and the Curie-Weiss interaction potential. These potentials possess different functional forms of the attractive component. The Morse potential contains a term that describes exponential decay of the attractive interaction with distance. The Curie-Weiss potential is given by a global non-local attractive term of equal strength for any pair of particles.

Analysis of data for short-range (Morse potential) and long-range (Curie-Weiss potential) attraction allows for a better understanding of the critical behavior pattern of systems with these two types of interaction. The analytical expressions used in numerical calculations and results are presented in Section~\ref{sec2} for the case of the Morse potential and in Section~\ref{sec3} for the case of Curie-Weiss-type interaction. Conclusions are given in Section~\ref{sec4}.

\section{Critical Point Parameters for Morse Fluids}
\label{sec2}

\subsection{General scope}
\label{sec2d1}

Our description of the critical behavior of fluids is carried out within the framework of the grand canonical ensemble (see, for example, \cite{hm113}) based on the cell fluid model proposed in \cite{kkd118,kkd120}.

We consider an open system of interacting particles in volume $V$ partitioned into $N_v$ cubic cells. Each cell has volume $v=V/N_v=c^3$, where $c$ is its linear size.

The Morse interaction potential as a function of distance $r\equiv|\ve{r}_{1} - \ve{r}_{2}|/R_0$ between particles with coordinates
$\ve{r}_{1},\,\ve{r}_{2}\in\mathbb R^3$ has the form \cite{morse29,kpd118,p120}
\bea
&&
\tilde U(r) = \Psi(r) - U(r); \non
&&
\Psi(r) = D e^{-2(r- 1)/\alpha_R}, \non
&&
U(r) = 2 D e^{-(r- 1)/\alpha_R}.
\label{1d1inf}
\eea
Here, $\Psi(r)$ and $U(r)$ denote the repulsive and attractive parts of the potential, respectively, and $\alpha_R = \alpha / R_0$, where $\alpha$ is the effective interaction radius. The parameter $R_0$ corresponds to the minimum of the function $\tilde U(r)$, and $D$ determines the depth of the potential well. We use $R_0$ as a reference length and measure the distances in units of $R_0$. Thus, the linear size of each cell $c$ and its volume $v$ are determined in units of $R_0$ and $R_0^3$, respectively.

In the renormalization group framework used in \cite{kpd118,pkd223}, the critical temperature for fluids is determied from the equation
\be
1 + q + R^{(0)} \sqrt{u^*} - \tilde a_2 \beta_c W(0) -
R^{(0)} \frac{a_4}{\sqrt{u^*}} (\beta_c W(0))^2 = 0,
\label{1d2inf}
\ee
where $\beta_c = (k_{\rm B} T_c)^{-1}$ is the inverse critical temperature, $k_{\rm B}$ is the Boltzmann constant, $T_c$ is the critical temperature. Further, $W(0)$ is the Fourier transform of the effective interaction potential at zero wave vector $\ve k$, and $\tilde a_2$, $a_4$ are coefficients in the initial expression for the grand partition function \cite{kpd118}. The quantity $q$ is related to averaging the square of the wave vector in the function $W(k)$. The value $R^{(0)}$ is determined by the elements and smallest eigenvalue of the renormalization group linear transformation matrix, and $u^*$ is the fixed-point coordinate.

At temperatures $T$ below the critical one, the reduced density  $\rho^* = (\langle N \rangle/V) R_0^3$, where $\langle N \rangle$ is the average number of particles in the system, and the effective chemical potential $M$ are related via the following nonlinear equation \cite{p120}
\be
\rho^* = \frac{R_0^3}{v} \left [ n_g - M + \sigma_{00}^{(-)}
\left ( \tilde h^2 + h^2_{cm} \right )^{\frac{d-2}{2(d+2)}} \right ].
\label{1d3inf}
\ee
Note, that in~\cite{p120}, this equation was written in terms of $\bar{n}=\langle N \rangle / N_v$, which is related to $\rho^*$ as $\bar{n} = \rho^* v / R_0^3$. The quantity $\tilde h$ is proportional to $M$, while $h_{cm}$ is determined by the relative temperature $\tau = (T-T_c)/T_c$. The first term $n_g$ on the right-hand side of (\ref{1d3inf}) is determined by the coefficients of the initial expression for the grand partition function. The coefficient $\sigma_{00}^{(-)}$ of the third term depends on the quantity $\alpha_m = \tilde h/h_{cm}$ depending both on the physical chemical potential $\mu$ through $M$, and the relative temperature $\tau$.
The space dimension $d$ is fixed at $d = 3$ in our current discussion.

The pressure $P$ as a function of temperature $T$ and density $\bar{n}$ is given by the equation of state \cite{kpd118,p120}. For temperatures below $T_c$, it is given by
\be
\frac{P v}{k_{\rm B} T} = P_a^{(-)}(T) + E_\mu +
\left ( \frac{\bar n - n_g}{\sigma_{00}^{(-)}} \right )^6
\Biggl[ e_0^{(-)} \frac{\alpha_m}{(1+\alpha_m^2)^{1/2}} +
\gamma_s^{(-)} - e_2^{(-)} \Biggr].
\label{1d4inf}
\ee
The quantity $P_a^{(-)}(T)$ contains an analytical temperature dependence. The coefficient $\gamma_s^{(-)}$ characterizes a nonanalytical contribution to the thermodynamic potential. The quantities $e_0^{(-)}$ and $e_2^{(-)}$, as well as $\sigma_{00}^{(-)}$, are related to the roots of an associated cubic equation.
Explicit expressions for all these quantities, as well as for $E_{\mu}$, are given in \cite{p120}.

\subsection{Numerical results}
\label{sec2d2}

We employed the above microscopic description of critical behavior of the Morse fluid to compute the critical values for the reduced temperature $T^*$, the reduced density $\rho^*$, and the reduced pressure $P^*$. These quantities are related to their dimensional counterparts via
\be
\frac{k_{\rm B} T}{D} = T^*, \quad
\rho \frac{N_{\rm A} R_0^3}{M_{mol}} = \rho^*, \quad
\frac{P R_0^3}{D} = P^*,
\label{1d7inf}
\ee
where $k_{\rm B} = 1.380649 \times 10^{-23}$ J $\cdot$ K$^{-1}$ is the Boltzmann constant,
$N_{\rm A} = 6.02214076 \times 10^{23}$ mol$^{-1}$ is the Avogadro number, and $M_{mol}$ is the molar mass.
The critical temperature $T_c^*$ is calculated using equation (\ref{1d2inf}). The critical density $\rho_c^* $
and critical pressure $P_c^* $ are obtained from (\ref{1d3inf}) and (\ref{1d4inf}), respectively,
at $M = 0$ and $T = T_c$.

We use the Morse interaction potential parameters characteristic for sodium (molar mass $M_{\rm Na}=22.9898$ g $\cdot$ mol$^{-1}$) and potassium (molar mass $M_{\rm K}=39.0983$ g $\cdot$ mol$^{-1}$).
These specific parameters are \cite{singh,lkg167}:
$R_0 = 5.3678$ \AA, $D = 0.9241 \times 10^{-13}$ erg for Na, and
$R_0 = 6.4130$ \AA, $D = 0.8530 \times 10^{-13}$ erg for K.
Our results for the critical point are \\[5pt]
--- for Na:
\bea
&&
T_c^* = 4.0277, \non
&&
\rho_c^* = 0.9971, \non
&&
P_c^* = 0.4735;
\label{1d5inf}
\eea
--- for K:
\bea
&&
T_c^* = 3.3043, \non
&&
\rho_c^* = 0.9347, \non
&&
P_c^* = 0.4075.
\label{1d6inf}
\eea

To facilitate comparison with the results of other studies, such as computer simulation \cite{singh} and experiment \cite{hensel}, we transform our dimensionless quantities from \eqref{1d5inf} and \eqref{1d6inf} into dimensional ones employing the relations \eqref{1d7inf}. Thus, we obtain numerical estimates of the critical point parameters for Na and K presented in Table~\ref{tab_1inf}.
\begin{table}[htbp]
\noindent\caption{Critical temperature $T_c$, density $\rho_c$, and pressure $P_c$ for sodium (Na) and potassium (K): cell fluid model (CFM) in the $\rho^4$ approximation vs Monte Carlo simulation and experimental results.}
\vskip3mm\tabcolsep4.5pt
\label{tab_1inf}

\begin{center}
\noindent{\footnotesize\begin{tabular}{|c|c|c|c|c|c|c|}
\hline%
\multicolumn{1}{|c}{\rule{0pt}{5mm}{Research method}} & \multicolumn{3}{|c}{Na} &
\multicolumn{3}{|c|}{K}\\[2mm]%
\cline {2-7}%
& \multicolumn{1}{|c}{\rule{0pt}{5mm}{$T_c$ (K)}} & \multicolumn{1}{|c}{$\rho_c$ (kg/m$^3$)} & \multicolumn{1}{|c}{$P_c$ (bar)}
& \multicolumn{1}{|c}{$T_c$ (K)} & \multicolumn{1}{|c}{$\rho_c$ (kg/m$^3$)} & \multicolumn{1}{|c|}{$P_c$ (bar)}\\[2mm]%
\hline%
\multicolumn{1}{|l|}{\rule{0pt}{5mm}{Theory\,(CFM,\,$\rho^4$\,model) \cite{kpd118,p120}}} & 2696 & 246 & 283 & 2042 & 230 & 132\\%
\multicolumn{1}{|l|}{Simulation \cite{singh}}
& 3932 & 353 & 1290 & 3120 & 277 & 534\\%
\multicolumn{1}{|l|}{Experiment \cite{hensel}} & 2485 & 300 & 248 & 2280 & 190 & 161\\[2mm]%
\hline
\end{tabular}}
\end{center}
\end{table}

\subsection{Behavior of critical point parameters with changes in the width of the attractive part of the Morse potential}
\label{sec2d3}

For potassium (K), the ratio $R_0/\alpha = 3.0564$ is somewhat larger than the analogous quantity $R_0/\alpha = 2.9544$ for sodium (Na) \cite{singh}. A larger $R_0/\alpha$ implies a smaller potential well width, meaning a narrower attractive part of the Morse potential.

From Table~\ref{tab_1inf}, we observe that both the critical temperature $T_c$ and critical pressure $P_c$ are smaller for K than for Na. This suggests that narrowing the attractive part of the Morse potential leads to smaller values of $T_c$ and $P_c$.
This conclusion aligns with the findings of \cite{okumura_00}, where liquid-vapor coexistence curves were computed using the $NPT$ plus test particle method for
various interatomic model potentials, including Lennard-Jones, 
Morse, and modified Stillinger-Weber.
Furthermore, our results indicate that the critical density $\rho_c$ also decreases  as the attractive part of the interaction potential narrows.

The liquid-gas critical point parameters for liquid alkali metals sodium and potassium, derived using the approach developed in \cite{kpd118,p120} and presented in Table \ref{tab_1inf}, show better agreement with experimental data \cite{hensel} than the Monte Carlo results from \cite{singh}. This is consistent with the authors' note in \cite{singh} that their critical pressure results for Na and K (quoted in Table \ref{tab_1inf}) deviate significantly from experimental values due to an overestimation of the critical temperature.

\section{Critical Point Parameters for Fluid Systems with the Curie-Weiss Potential}
\label{sec3}

Let us turn our attention to a one-component continuous system with Curie-Weiss-type interactions. We begin by presenting some basic formulas from \cite{kkd120} and \cite{kd122} used in this investigation.

\subsection{Key definitions and relations}
\label{sec3d1}

The Curie-Weiss-type interaction potential is given by
\begin{equation}
	\Phi_{N_v} (\ve{r}, \ve{r}') = - \frac{g_a}{N_v} + g_r \sum_{l =1}^{N_v} I_{\Delta_\ell}(\ve{r}) I_{\Delta_\ell}(\ve{r}'),
	\label{2d1inf}
\end{equation}
where
\begin{equation}
		I_{\Delta_\ell}(\ve{r}) = \left\{
		\begin{array}{ll}
			1, & \ve{r}\in \Delta_\ell \\
			0, & \ve{r}\notin \Delta_\ell
		\end{array}
		\right.
	\label{2d2inf}
\end{equation}
is the indicator function of the cell $\Delta_\ell$, for $l=1,\ldots,N_v$. As in Subsection~2.1, we consider a system of interacting particles in a volume $V$, which comprises $N_v$ congruent cells,
each of volume $v=c^3$.

The term with $g_a>0$ describes a uniform attraction influencing all particle pairs in the system, while the term with $g_r>0$ is responsible for a short-range repulsion between two particles within the same cell. The stability of the interaction \cite{r170} is ensured by the inequality $g_r>g_a>0$.
\label{pcon}

Previous studies \cite{kkd120,kd122} have established that the cell model under consideration admits an exact representation in the form of a single integral, whose asymptotic value in the thermodynamic limit is determined via Laplace's method.
To achieve this, we need to find the maximum value $E_0(\bar z, \mu)$ of the function
\be
E_0(z, \mu) = - (z - \beta\mu-\ln v^*)^2 / 2 p_a + \ln K_0(z),
\label{2d5inf}
\ee
where $z = \bar z$ denotes the location of the global maximum.

The special function $K_0(z)$ is a representative of the family of functions
\be
K_m(z) = \sum_{n=0}^{\infty} \frac{n^m}{n!}
\exp\left(-\frac{fp_a}{2}n^2\right)\exp(z n), \quad
m = 0,1,2,\ldots.
\label{2d6inf}
\ee
These functions depend on the repulsion strength $g_r$ and temperature through the definitions
\be
f = g_r/g_a, \quad fp_a = p_r, \quad p_r = \beta g_r, \quad \beta = (k_{\rm B} T)^{-1}.
\label{2d7inf}
\ee
The dimensionless volume $v^*=v/\Lambda^3$ is considered equal to $v$, as the de Broglie wavelength $\Lambda$ is set to $1$.

The chemical potential $\mu$ is given as a function of $\bar z$ by the relation
\be
\beta\mu = \bar z - p_a \frac{K_1(\bar z)}{K_0(\bar z)}  - \ln v^*,
\label{2d3inf}
\ee
where $p_a=\beta g_a$.

For the average particle density we have
\be
\bar n = \frac{K_1(\bar z)}{K_0(\bar z)}.
\label{2d8inf}
\ee

The pressure $P$ is obtained from the equation of state
\begin{equation}
P v \beta_c = (\tau + 1) \left[\ln K_0(\bar z) -
 \frac{p_a}{2}\left(\frac{K_1(\bar z)}{K_0(\bar z)}\right)^2\right].
\label{2d10inf}
\end{equation}

The Curie-Weiss cell model exhibits a sequence of first-order phase transitions, each terminating at a critical point with a specific critical temperature \cite{kd122}. In terms of the model parameters introduced just above, each critical temperature associated with the $n$-th critical point is given by
\begin{equation}
k_B T_c^{(n)} = g_a / p_{ac}^{(n)}.
\label{2d9inf}
\end{equation}
Thus, the superscript $n$ denotes the sequential number of the critical point in the series, starting from the \emph{lowest} critical temperature $T_c^{(1)}$. The specific values of the critical temperatures $T_c^{(n)}$ depend on the chosen value of the Curie-Weiss attraction strength $g_a$. Since we are presently interested in the general behavior of the system, there is no need to assign a specific value to $g_a$. To simplify notation, we assume
\be\label{B1}
T_c^{(1)}\equiv T_c\quad\mbox{and}\quad\beta_c^{(1)}\equiv\beta_c.
\ee

Table~\ref{tab_2inf} presents our numerical results obtained for the first three critical points in the sequence, with the lowest critical temperatures $T_c^{(n)}$. The (arbitrary) cell volume $v^*$ is set to $v^*=5.0$. The interaction-strength ratio $f=g_r/g_a$ ranges from $f\gtrsim1$ close to the marginal value $f=1$ dictated by the stability condition $g_r>g_a>0$ (see p.~\pageref{pcon}) to $f=10$ corresponding to the asymptotic regime $f\gg1$. The specific case $f=1$ is elaborated in \cite{DSh24}. Further discussion is deferred to the following section.
\begin{table}[htbp]
\noindent\caption{Critical parameters for the first three critical points with the lowest critical
temperatures $T_c^{(n)}=1/p_{ac}^{(n)}$ at different values of $f=g_r/g_a$ and a fixed $v^*=5.0$.
The value $\beta_c$ refers to $n=1$ as in \eqref{B1}.}
\vskip3mm\tabcolsep4.5pt
\label{tab_2inf}

\begin{center}
\noindent{\footnotesize\begin{tabular}{|c|c|c|c|c|c|}
\hline%
{\rule{0pt}{5mm}$f$} & $n$ & $T_c^{(n)}$ & $\bar n_c^{(n)}$ &
$P_c^{(n)}v\beta_c$ & $\beta_c\mu_c^{(n)}$\\[2mm]%
\hline%
\rule{0pt}{5mm} & 1 & $0.2614$ & 0.5355 & 0.2048 & $-1.6405$\\%
 1.00000001 & 2 & 0.2841 & 1.5191 & 1.0261 & $-1.0098$\\%
 & 3 & 0.2988 & 2.5128 & 2.0406 & $-0.5921$\\%
\hline
 & 1 & $0.2614$ & 0.5355 & 0.2048 & $-1.6405$\\%
 1.0000001 & 2 & 0.2841 & 1.5191 & 1.0261 & $-1.0098$\\%
 & 3 & 0.2988 & 2.5128 & 2.0406 & $-0.5921$\\%
\hline
 & 1 & $0.2614$ & 0.5355 & 0.2048 & $-1.6405$\\%
 1.000001 & 2 & 0.2841 & 1.5191 & 1.0261 & $-1.0098$\\%
 & 3 & 0.2988 & 2.5128 & 2.0406 & $-0.5921$\\%
\hline
 & 1 & $0.2614$ & 0.5355 & 0.2048 & $-1.6405$\\%
 1.00001 & 2 & 0.2841 & 1.5191 & 1.0262 & $-1.0097$\\%
 & 3 & 0.2988 & 2.5128 & 2.0407 & $-0.5920$\\%
\hline
 & 1 & $0.2614$ & 0.5355 & 0.2048 & $-1.6403$\\%
 1.0001 & 2 & 0.2841 & 1.5191 & 1.0264 & $-1.0092$\\%
 & 3 & 0.2987 & 2.5128 & 2.0415 & $-0.5911$\\%
\hline
 & 1 & $0.2614$ & 0.5353 & 0.2047 & $-1.6384$\\%
 1.001 & 2 & 0.2840 & 1.5190 & 1.0291 & $-1.0035$\\%
 & 3 & 0.2985 & 2.5127 & 2.0495 & $-0.5819$\\%
\hline
 & 1 & $0.2609$ & 0.5338 & 0.2042 & $-1.6200$\\%
 1.01 & 2 & 0.2822 & 1.5177 & 1.0562 & $-0.9467$\\%
 & 3 & 0.2955 & 2.5116 & 2.1314 & $-0.4905$\\%
\hline
 & 1 & $0.2590$ & 0.5277 & 0.2023 & $-1.5379$\\%
 1.05 & 2 & 0.2756 & 1.5134 & 1.1834 & $-0.6960$\\%
 & 3 & 0.2850 & 2.5082 & 2.5191 & $-0.0855$\\%
 \hline
 & 1 & $0.2571$ & 0.5219 & 0.2004 & $-1.4351$\\%
 1.1 & 2 & 0.2695 & 1.5097 & 1.3541 & $-0.3854$\\%
 & 3 & 0.2760 & 2.5056 & 3.0420 & 0.4195\\%
\hline
 & 1 & $0.2546$ & 0.5139 & 0.1977 & $-1.2298$\\%
 1.2 & 2 & 0.2619 & 1.5056 & 1.7194 & 0.2309\\%
 & 3 & 0.2653 & 2.5030 & 4.1567 & 1.4292\\%
\hline
 & 1 & $0.2530$ & 0.5090 & 0.1961 & $-1.0252$\\%
 1.3 & 2 & 0.2575 & 1.5034 & 2.1018 & 0.8432\\%
 & 3 & 0.2594 & 2.5018 & 5.3174 & 2.4394\\%
\hline
 & 1 & $0.2513$ & 0.5039 & 0.1944 & $-0.6184$\\%
 1.5 & 2 & 0.2531 & 1.5014 & 2.8883 & 2.0604\\%
 & 3 & 0.2539 & 2.5007 & 7.6917 & 4.4578\\%
\hline
 & 1 & $0.2506$ & 0.5017 & 0.1937 & $-0.2143$\\%
 1.7 & 2 & 0.2514 & 1.5006 & 3.6852 & 3.2708\\%
 & 3 & 0.2517 & 2.5003 & 10.0889 & 6.4708\\%
\hline
 & 1 & $0.2502$ & 0.5005 & 0.1933 & 0.3887\\%
 2.0 & 2 & 0.2504 & 1.5002 & 4.8851 & 5.0787\\%
 & 3 & 0.2505 & 2.5001 & 13.6923 & 9.4817\\[2mm]%
\hline
\end{tabular}}
\end{center}
\end{table}
\begin{table}[htbp]
{\footnotesize\textit{Table~\ref{tab_2inf},} \textbf{continued}.}\vskip0.7mm\tabcolsep4.5pt

\begin{center}
\noindent{\footnotesize\begin{tabular}{|c|c|c|c|c|c|}
\hline%
{\rule{0pt}{5mm}$f$} & $n$ & $T_c^{(n)}$ & $\bar n_c^{(n)}$ &
$P_c^{(n)}v\beta_c$ & $\beta_c\mu_c^{(n)}$\\[2mm]%
\hline%
\rule{0pt}{5mm} & 1 & $0.2500$ & 0.5001 & 0.1932 & 1.3902\\%
 2.5 & 2 & 0.2500 & 1.5000 & 6.8860 & 8.0828\\%
 & 3 & 0.2501 & 2.5000 & 19.6960 & 14.4877\\%
 \hline
 & 1 & $0.2500$ & 0.5000 & 0.1932 & 2.3905\\%
 3.0 & 2 & 0.2500 & 1.5000 & 8.8862 & 11.0835\\%
 & 3 & 0.2500 & 2.5000 & 25.6970 & 19.4889\\%
 \hline
 & 1 & $0.2500$ & 0.5000 & 0.1931 & 3.3906\\%
 3.5 & 2 & 0.2500 & 1.5000 & 10.8863 & 14.0837\\%
 & 3 & 0.2500 & 2.5000 & 31.6972 & 24.4891\\%
 \hline
 & 1 & $0.2500$ & 0.5000 & 0.1931 & 4.3906\\%
 4.0 & 2 & 0.2500 & 1.5000 & 12.8863 & 17.0837\\%
 & 3 & 0.2500 & 2.5000 & 37.6972 & 29.4892\\%
\hline
 & 1 & $0.2500$ & 0.5000 & 0.1931 & 6.3906\\%
 5.0 & 2 & 0.2500 & 1.5000 & 16.8863 & 23.0837\\%
 & 3 & 0.2500 & 2.5000 & 49.6972 & 39.4892\\%
 \hline
 & 1 & $0.2500$ & 0.5000 & 0.1931 & 10.3906\\%
 7.0 & 2 & 0.2500 & 1.5000 & 24.8863 & 35.0837\\%
 & 3 & 0.2500 & 2.5000 & 73.6972 & 59.4892\\%
\hline
 & 1 & $0.2500$ & 0.5000 & 0.1931 & 16.3906\\%
 10.0 & 2 & 0.2500 & 1.5000 & 36.8863 & 53.0837\\%
 & 3 & 0.2500 & 2.5000 & 109.6972 & 89.4892\\[2mm]%
\hline
\end{tabular}}
\end{center}
\end{table}

\subsection{Influence of the attraction strength on critical parameters}
\label{sec3d2}

This section analyzes the critical parameters of our cell model as presented in Table~\ref{tab_2inf}. We consider the critical values of pressure, temperature, density, and chemical potential. Our numerical results complement and extend the findings of \cite{dkpppr125}.

\begin{figure}[htbp]
\vskip1mm
\centering \includegraphics[width=0.65\textwidth]{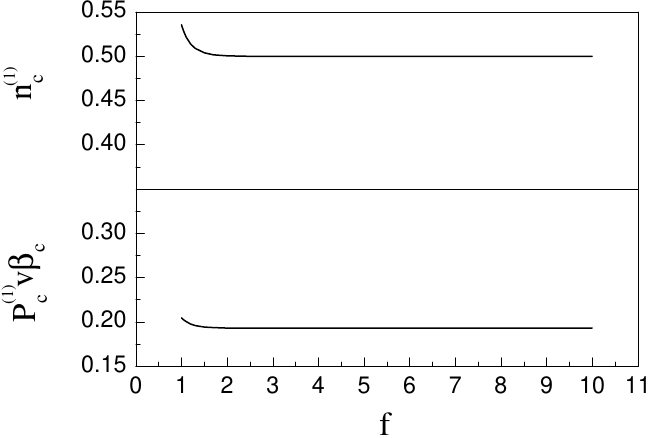}
\vskip-3mm
\caption{Critical density $\bar n_c^{(1)}$ and critical pressure $P_c^{(1)}v\beta_c$ for the first critical point as a function of parameter $f$ (with $v^*=5.0$).}
\label{fig_1inf}
\end{figure}

\begin{figure}[htbp]
\vskip1mm
\centering \includegraphics[width=0.65\textwidth]{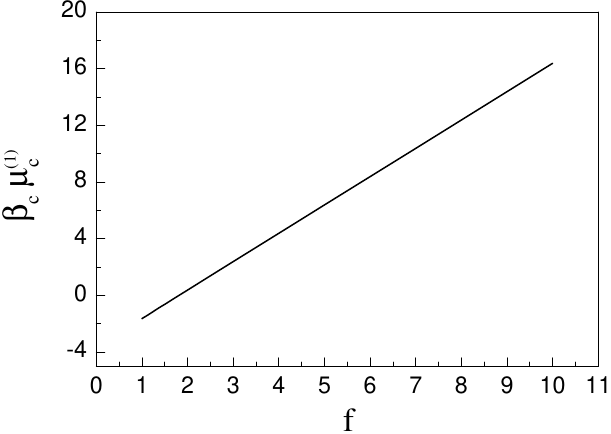}
\vskip-3mm
\caption{Change in the critical chemical potential $\beta_c\mu_c^{(1)}$ for the first phase transition as a function of parameter $f$ (with $v^*=5.0$).}
\label{fig_2inf}
\end{figure}

The parameter $f = g_r/g_a$, introduced in \eqref{2d7inf}, is inversely proportional to the attraction strength $g_a$ and must satisfy the stability condition $f > 1$. Therefore, while fixing the repulsion strength $g_r$, we observe in Table~\ref{tab_2inf} how the critical parameters change as the global Curie-Weiss attraction $g_a$ decreases.
\begin{enumerate}
\item
 For each specific value of $f$, the critical temperature, critical density, pressure, and chemical potential all increase when transitioning to each subsequent critical point in the sequence. This increase, however, essentially slows down for the critical temperatures $T_c^{(n)}$. Specifically, for large $f$, such as $f > 3.0$, where repulsion significantly dominates attraction, the critical temperature for all first three phase transitions asymptotically approaches the same value $T_c$. In this case, $g_a/k_B T_c^{(n)} = p_{ac}^{(n)} \rightarrow 4.0$. This behavior is in agreement with analytical calculations performed in \cite[Sec. 7]{DSh24}. This reference also contains a detailed treatment of another marginal case, $f=1$.
\item
We now analyze the behavior of the critical point parameters corresponding to the same index $n$ for varying values of $f$. When $f \rightarrow 1$, all critical parameters for a given $n$ tend to their limiting values (see Table~\ref{tab_2inf}).
As the parameter $f$ increases (corresponding to a decrease in attraction strength $g_a$), the reduced critical pressure $P_c^{(1)}v\beta_c$ shows a negligible decrease, while the pressures $P_c^{(n)}v\beta_c$ for subsequent critical points with $n=2$ and $n=3$ increase significantly.
The critical temperatures $T_c^{(n)}$ gradually decrease, eventually becoming identical for large $f$ values as discussed in the previous entry.
The critical density $\bar n_c^{(n)}$  also decreases negligibly with increasing $f$. Notably, regardless of the repulsion-to-attraction ratio $f$, the critical points consistently appear at approximately the same mean densities: $0.5$, $1.5$, and $2.5$.
The slight decrease in critical density and pressure for the first critical point and their subsequent saturation for large $f$ are illustrated in Fig.~\ref{fig_1inf}. The critical chemical potential values increase with $f$, and a typical picture is given in Fig.~\ref{fig_2inf} for $n=1$.
\end{enumerate}

\section{Conclusions}
\label{sec4}

A characteristic feature of this study is the use of microscopic information about the interparticle interaction potential to evaluate the critical point parameters.

In the first part of the paper, we investigate the influence of the attractive part of the interaction potential on the liquid-gas critical point in Morse fluids. Numerical estimates are obtained using Morse potential parameters characteristic of alkali metals, specifically sodium and potassium. We converted the theoretical dimensionless critical point parameters into dimensional quantities often used in numerical and experimental studies.
We show that a decrease in the width of the attractive part of the Morse potential leads to a reduction in both the critical temperature and pressure for both metals.
Our results are in good agreement with existing data from other sources.

For the fluid cell model analyzed in the second part of this work, we investigated how the interparticle Curie-Weiss-type attraction strength influences the parameters for the first three critical points.
We accomplished this by studying results obtained with varying ratios $f = g_r/g_a$ of repulsion and attraction strenths.

For a specific value of $f$, the critical point parameters increase upon transitioning to each subsequent $n$. When attraction is significantly weaker than repulsion ($f > 3.0$), the critical temperature for all first three phase transitions becomes identical.

For varying $f$, we examined the critical points with the same index $n$, see Table~\ref{tab_2inf}. Thus, we found that as the ratio $f$ increases (i.e., attraction strength decreases), the critical temperatures $T_c^{(n)}$ gradually decrease, eventually becoming identical for large $f$.
At the same time, we observed a slight decrease in critical density and pressure for the first critical point, which then saturated. The critical density for $n=2$ and $n=3$ decrease negligibly with increasing $f$. Regardless of $f$, the critical points consistently appeared at approximately the same mean densities. An increase in $f$ also led to higher critical chemical potential in all considered cases.

Let us compare the influence of the attractive parts of the Morse and Curie-Weiss
potentials on the liquid-gas critical point. The attractive components of these
potentials have different functional forms. As mentioned above, narrowing the width
of the attractive part of the Morse potential decreases the values of
both the critical temperature and pressure. For the Curie-Weiss potential, a decrease
in the attraction strenght $g_a$ (or an increase in the parameter $f = g_r/g_a$)
leads to an increase in the quantity $p_{ac}^{(1)}$ in the expression \eqref{2d9inf}
for the critical temperature \cite{dkpppr125}, and therefore, to a decrease
in $T_c^{(1)}\equiv T_c$. The critical pressure for the first phase transition
point also decreases with an increase in $f$ (see Table~\ref{tab_2inf}, $n = 1$).
Thus, a consistent correlation is observed in how the attractive parts of two
potentials of different types affect the behavior of the temperature and pressure
at the liquid-gas critical point.

\vskip3mm \textit{
This work was supported by the National Research Foundation of Ukraine under
	the project No. 2023.03/0201.\\
The authors are deeply grateful to all warriors of the Ukrainian Armed Forces, living and fallen, for making this research possible.}

\end{document}